\documentclass[seceq]{ptptex}

\usepackage{graphicx}

\notypesetlogo                       

\title{
The von Neumann Entropy of EPR Spin Correlation for the Relativistic Pairs 
}

\author{
Yoshihisa \textsc{Nishikawa}%
}

\inst{
Department of Physics, Konan University, 8-9-1 Okamoto, Kobe 658-8501, Japan
}

\abst{
Variation of the von Neumann entropy by the Lorentz transformation is discussed. Taking the spin-singlet state in the center of mass frame, the von Neumann entropy in the laboratory frame is calculated from the reduced density matrix obtained by taking the trace over 4-momentum after the Lorentz transformation.
As the model to discuss the EPR spin correlation, it is supposed that one parent particle splits into a superposition state of various pair states in various directions. Computing the von Neumann entropy and the Shannon entropy, we have shown a global behavior of the entropy to see a relativistic effect. We discuss also the super-relativistic limit, distinguishability between the two particles of the pair and so on.
}

\begin{document}

\maketitle


\section{Introduction}

Recently, quantum states are treated as a resource of the quantum information.\cite{Gruska,QCAQI} \ In the information theory, the Shannon entropy quantifies the amount of information and the von Neumann entropy in quantum mechanics quantifies the purity of state.\cite{PERES}

In around 2000, it is recognized in a study of the quantum information that spin states are affected by the nontrivial relativistic effect due to the the successive Lorentz transformation.\cite{PeresScudoTerno}\tocite{HeShaoZhang02} \ The von Neumann entropy for a single particle has been shown to be not Lorentz invariant.\cite{PeresScudoTerno} \ It has been discussed that the degree of the violation of Bell's inequality decreases in the relativistic regime.\cite{Czachor1997}\tocite{Ahn2} \ Entropy is also used to estimate some EPR correlations taking into account the relativistic effect.\cite{HeShaoZhang01,HeShaoZhang02}

Since Einstein, Podolsky and Rosen (EPR), \cite{EPR} \ the quantum correlation called an entanglement among separate systems has been discussed in various models. After Bohm formulated this discussion in terms of spin correlation,\cite{BOHM} many people have discussed this correlation in such model that one parent particle splits into a pair of particles in the spin-singlet state running away in the opposite directions. The hypotheses of hidden variable had been discussed assuming such situation. Assuming only the local causality, Bell derived an inequality formula\cite{BELL,ASPECT} \ to be satisfied for the statistical data of the separate systems which had been in contact prior to the separation. 

The parent particle which splits into the pair will be generally running in the laboratory frame. Threfore, we introduce the center of mass frame in which the parent particle is at rest. 
In this center of mass frame, there are two spin-$\frac{1}{2}$ particles with same mass and same absolute magnitude of 3-momentum. Then the state vector of the pair is assumed to be a factor state of spin part and 4-momentum part. For such factor state, the von Neumann entropy is zero in general. By applying the Lorentz transformation into the laboratory frame, the state vector can be changed into an entangled state between the spin part and the 4-momentum part. Therefore, the von Neumann entropy calculated from the reduced density matrix, which is obtained by taking the trace over 4-momentum in the laboratory frame, has now a finite value in general. 

In this paper, we discuss the spin correlation in an extended model where the state vector after the splitting is a superposition state of various pair states, instead of the pair in one direction as in the previous models. The various pair states are classified by various velocity directions of each pair. This model is envisaged as a simplification of such physical situation that the ejection direction of the pair from the parent particle, nucleus or elementary particle, spreads in a finite range of angle. In our model, however, the angular distribution is discretized and it is represented by the superposition state of various pair states with a definite angle of ejection. In this paper, we treat only the two pair states as the simplest case of the extended model. 

Using the reduced density matrix for the extended model of various pair states, the von Neumann entropy is calculated and we discuss the correlation among the composite spin states defined with respect to the $z$-direction in the space-fixed frame. The value of the entropy is considered to describe how much degree the spin correlation, which is assigned in the center of mass frame, is degraded in the laboratory frame. Since the spin state is coupled with 4-momentum in the Lorentz transformation, the computed values of the entropy are dependent on the ejection velocity and the velocity of the center of mass frame relative to the laboratory frame. It is shown in this paper that the entropy does work as a measure to quantify the spin correlation in this extended model even for the relativistic case.

In section \ref{Lorentz transformation of state vector and Wigner rotation}, the unitary transformation of the state vector induced by the Lorentz transformation is discussed. Due to the Wigner rotation resulted from the successive Lorentz transformation,\cite{WEINBERG} the spin state becomes to contain the triplet-state in the laboratory frame even if the singlet-state is prepared in the center of mass frame.
In section \ref{An extension of the splitting model of EPR}, we propose an extended model to consider the EPR spin correlation, where the state vector after the splitting is a superposition of various pair states. In section \ref{Transformation of spin-singlet state by the Wigner rotation}, the mixing coefficients of the spin state due to the Wigner rotation is calculated for the simplest case of two pair states. 

In section \ref{The von Neumann entropy of spin correlation in the laboratory frame}, we calculate the von Neumann entropy and show the results in figures. In section \ref{The von Neumann entropy and the Shannon entropy}, by comparing the Shannon entropy and the von Neumann entropy, the effect of distinguishability between the two particles is discussed. In section \ref{Discussion}, other features are discussed. We use the base-2 of logarithm and natural unit: $\hbar=c=1$. 

\section{Lorentz transformation of state vector and Wigner rotation}\label{Lorentz transformation of state vector and Wigner rotation}

We consider a state vector of a massive particle with 4-momentum $p^\mu $ and spin $\sigma$ in a coordinate frame. 
\begin{eqnarray}
P^\mu |p,\sigma \rangle &=&p^\mu |p,\sigma \rangle , \label{p,sigma}\\
S_z |k,\sigma\rangle 	&=&\sigma |k,\sigma\rangle . \label{spin defined with the rest 4-momentum}
\end{eqnarray}
Here, $P^\mu $ is the 4-momentum operator, $S_z$ is the $z$ component of spin operator and $k^\mu =\{m,0,0,0\}$ is the rest 4-momentum of the particle with mass $m\ne 0$. The state $|p,\sigma \rangle $ is connected with $|k,\sigma \rangle $ by a unitary operator $U(L(p))$ corresponding to a boost transformation $L(p)$ as follows:
\begin{eqnarray}
|p,\sigma \rangle =N(p)U(L(p))|k,\sigma \rangle , \label{k->p}
\end{eqnarray}
where $N(p)$ is a normalization factor.

Applying another boost transformation of $\Lambda $ on $|p,\sigma \rangle $, we get the state vector in the frame boosted by $\Lambda $, in which the 4-momentum is $\Lambda p$, as follows:
\begin{eqnarray}
U(\Lambda)|p,\sigma \rangle 	&=&	U(\Lambda)[N(p)U(L(p))|k,\sigma \rangle ] \nonumber \\
				&=&	N(p)U(\Lambda L(p))|k,\sigma \rangle 	\nonumber \\
				&=&	N(p)[U(L(\Lambda p))U^{-1}(L(\Lambda p))]U(\Lambda L(p))|k,\sigma \rangle \nonumber \\
				&=&	N(p)U(L(\Lambda p))U(W(\Lambda ,p))|k,\sigma \rangle ,
\label{p_acted}
\end{eqnarray}
where 
\begin{eqnarray}
W(\Lambda ,p)=L^{-1}(\Lambda p)\Lambda L(p) ~~\text{and}~~\Lambda p=\Lambda ^\mu {}_\nu p^\nu \label{wigner_rotation}.
\end{eqnarray}
In case of a massive particle, $W(\Lambda ,p)$ represents a spatial rotation and it is called the Wigner rotation. In contrast to the Galilei transformation, the Lorentz transformation affects the time-component of 4-momentum. Therefore, a non-trivial effect such as the Wigner rotation of Eq. (\ref{wigner_rotation}) is induced by a successive boost transformation in the different directions. Thus, spin component coupled with the 4-momentum is affected. 

The Eq. (\ref{p_acted}) can be written in reference to the spin state in the rest frame of particle as 
\begin{eqnarray}
U(\Lambda)|p,\sigma \rangle 	&=&	N(p)U(L(\Lambda p))\sum_{\sigma '}|k,\sigma '\rangle \langle k,\sigma'|U(W(\Lambda ,p))|k,\sigma \rangle  \nonumber \\
				&=&	N(p)\sum_{\sigma '} \langle k,\sigma'|U(W(\Lambda ,p))|k,\sigma \rangle U(L(\Lambda p))|k,\sigma '\rangle \nonumber \\
				&=&	\frac{N(p)}{N(\Lambda ,p)}\sum _{\sigma '}D_{\sigma ', \sigma }(W(\Lambda ,p))|\Lambda p,\sigma ' \rangle ,
\label{p->Lambda_p}
\end{eqnarray}
where
\begin{eqnarray}
D_{\sigma '\sigma }(W(\Lambda ,p))=\langle k,\sigma'|U(W(\Lambda ,p))|k,\sigma \rangle.
\label{wigner rotation D}
\end{eqnarray}
Therefore, even if a particle is in an eigenstate of spin in some frame, the spin state becomes a superposition state among other eigenstates of spin in the boosted frame. 

In the following discussion, we omit the normalization factors such as $N(p)$, $N(\Lambda ,p)$ and $\frac{N(p)}{N(\Lambda ,p)}$ because these do not affect the result of our discussion.

\section{An extension of the splitting model of EPR}\label{An extension of the splitting model of EPR}

For a long time, the EPR paradox had been told in such a model, in which one parent particle splits into one pair of particles. In this paper, we extend this old model into the model in which the state after the splitting is a superposition state of various pair states instead of the one pair state. 
We discuss the spin correlation of this superposition state. In the coordinate frame I, a composite spin state for each pair state is supposed to be in the spin-singlet state as follows: 
\begin{eqnarray}
|\varphi \rangle =
	\frac{1}{\sqrt{2}}
	\left(
		|p,\uparrow \rangle |q,\downarrow \rangle 
		-|p,\downarrow \rangle |q,\uparrow \rangle 
	\right),
	\label{singlet state with momentum}
\end{eqnarray}
where $p$ and $q$ are the 4-momentums of the particles. The spin-singlet state is an example of quantum entanglement.

\begin{figure}[htbp]
		\centerline{\includegraphics[width=6.6cm]{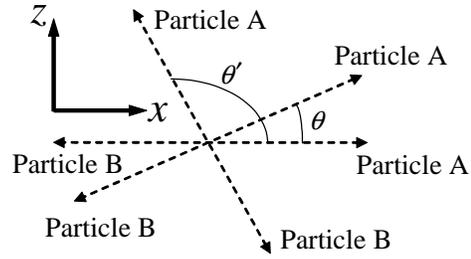}}
	\caption{The case of $i=1, 2, 3$. }
	\label{a figure of the extended model}
\end{figure}

We now consider the various spin-singlet pair states $|\varphi _i \rangle $ labeled by $i=1,2,\cdots$. And the 4-momentum set of each pair state is denoted by $\{p, q\}_i$. We assume in this paper the masses of particle are identical for all pair states. Therefore, one particle's 4-momentum is represented by an absolute magnitude $p$ and a direction $\hat{p}$ of the 3-momentum; then, $\{p,q\}_i$ is written as $\{(p_0, p \hat{p}),(p_0, -p \hat{p})\}_i$. Using these states, we can write the superposition state of various pair states as 
\begin{eqnarray}
|\psi \rangle =\sum _i c_i |\varphi_i \rangle \:\text{with}\:\sum _i|c_i|^2=1. 
\end{eqnarray}
The old model corresponds to $|\psi \rangle =|\varphi \rangle $. Figure \ref{a figure of the extended model} illustrates the case of $i=$1, 2, 3.

In the following of this paper, our discussion is restricted mainly to two pair states, the case of $i=1$ and $2$, as the simplest extension. Furthermore, we consider that the orbits of the two particles lie on the $x$-$z$ plane. And then we take the other coordinate frame II boosted in $z$-direction. We have assumed such situation since the effect of the Wigner rotation is emphasized. The direction of pair-$i$ is represented by $\theta_i$, which is the angle from $x$-axis toward $z$-axis. The state vector $|\varphi _i \rangle $ is now written as
\begin{eqnarray}
|\varphi_i \rangle = |\varphi _p(\theta _i)\rangle =\frac{1}{\sqrt{2}}
	\left(
	|p(\theta _i),\uparrow \rangle |p(\theta _i+\pi),\downarrow \rangle 
	-|p(\theta _i),\downarrow \rangle |p(\theta _i+\pi),\uparrow \rangle 
	\right) . \label{spin-singlet state in the frame A}
\end{eqnarray}
The superposition state of two pair states after the splitting is written as
\begin{eqnarray}
|\psi _p(\theta_1,\theta_2)\rangle =
	c_1|\varphi _p(\theta_1)\rangle +c_2|\varphi _p(\theta_2)\rangle. 
	\label{superposed spin-singlet in the frame A}
\end{eqnarray}

We now consider the state vector in the coordinate frame II boosted by $\Lambda$ relative to the coordinate frame I. The coefficients of the Wigner rotation in Eq. (\ref{wigner rotation D}) depend not only on $\Lambda$ but also on $p$. Using the unitary operator $U(\Lambda )$, the state vector in the coordinate frame II is given as
\begin{eqnarray}
U(\Lambda)|\psi _p(\theta_1,\theta_2)\rangle =
	c_1U(\Lambda)|\varphi _p(\theta_1)\rangle +c_2U(\Lambda)|\varphi _p(\theta_2)\rangle .
	\label{superposed spin-singlet in the frame B}
\end{eqnarray}

We can suppose such a physical situation corresponding to this extended model: in the laboratory frame identified with the coordinate frame II, a parent particle has been moving in the minus $z$-direction and it has splited into the superposition state of $\theta _1$-pair and $\theta _2$-pair states. The relativistic effect for angular distribution is expected to be described even in this discretized simplification. We identify the center of mass frame with the coordinate frame I and this coordinate frame is obtained applying the Lorentz transformation $\Lambda ^{-1}$ in the $z$-direction of the laboratory frame. The state vector in the laboratory frame is represented by Eq. (\ref{superposed spin-singlet in the frame B}) and the state vector in the center of mass frame is represented by Eq. (\ref{spin-singlet state in the frame A}).

Note that the $z$-axis of the rest frame for each particle is inclined in the laboratory frame, even if the ejection velocity and the boost velocity are taken on the $x$-$z$ plane.

\section{Transformation of spin-singlet state by the Wigner rotation}\label{Transformation of spin-singlet state by the Wigner rotation}

We calculate the transformation coefficients of the spin state in the laboratory frame by the Wigner rotation for the following state in the center of mass frame:
\begin{eqnarray}
|\varphi _p(\theta )\rangle =\frac{1}{\sqrt{2}}
	\left(
	|p(\theta ),\uparrow \rangle |p(\theta +\pi),\downarrow \rangle 
	-|p(\theta ),\downarrow \rangle |p(\theta +\pi),\uparrow \rangle 
	\right) ,
	\label{spin_singlet_at_center_of_mass}
\end{eqnarray}
where $p(\theta)=\{p^0,p\cos \theta ,0,p\sin \theta \}$ and $p(\theta+\pi)=\{p^0,-p\cos \theta ,0,-p\sin \theta \}$. The up-down arrow expresses the $z$-component of spin in the center of mass frame as well as in the laboratory frame. 

We now consider the state vector in the laboratory frame by Eq. (\ref{superposed spin-singlet in the frame B}). We use the rapidity representation of velocity: the ejection velocity $v$ of massive particles as $\phi =\tanh^{-1}v$ and the velocity $V$ of the boost transformation as $\alpha =\tanh^{-1}V$. The rapidity parameters can be limited in the region of $\alpha$, $\phi > 0$ for our discussion. The Wigner rotation matrix is generally written as follows:\cite{Ahn}
\begin{eqnarray}
D(W(\Lambda ,p))&=&
\frac{1}{[\frac{1}{2}+\frac{1}{2}\cosh\alpha \cosh\phi +\frac{1}{2}\sinh\alpha \sinh\phi (\hat{e}\cdot \hat{p})]^\frac{1}{2}} \nonumber \\
	&\times &
		\left\{
		[\cosh\frac{\alpha }{2}\cosh\frac{\phi}{2}+\sinh\frac{\alpha }{2}\sinh\frac{\phi}{2}(\hat{e}\cdot \hat{p}) ]I
	+
		i \sinh\frac{\alpha }{2}\sinh\frac{\phi}{2}(\vec{\sigma }\cdot \hat{n})
		\right\}, \nonumber \\
\end{eqnarray}
where $i$ is the imaginary unit, $I$ is the identity matrix, $\vec{\sigma }=(\sigma _x,\sigma_y,\sigma _z)$ is the Pauli matrices, $\hat{p}$ is the direction of the particle in the center of mass frame and $\hat{n}$ is defined as $\hat{n}\equiv \hat{e}\times \hat{p}$, $\hat{e}$ being the moving direction of the center of mass frame in the laboratory frame. 

Since $\hat{p}=(\cos \theta ,0,\sin \theta )$ and $\hat{e}=(0,0,-1)$ in our specific model, the matrix representation of the Wigner rotation becomes as
\begin{eqnarray}
D(W(\phi ,\alpha ,\theta ))
=
\left(
\begin{array}{cc}
 	A(\phi ,\alpha ,\theta )& 	-B(\phi ,\alpha ,\theta ) \\
 	B(\phi ,\alpha ,\theta )& 	A(\phi ,\alpha ,\theta )
\end{array}
\right), \label{wigner matrix}
\end{eqnarray}
where
\begin{eqnarray}
A(\phi ,\alpha ,\theta )&=&\frac{
\cosh\frac{\alpha }{2}\cosh\frac{\phi}{2}-\sinh\frac{\alpha }{2}\sinh\frac{\phi}{2}\sin \theta 
}{
[\frac{1}{2}+\frac{1}{2}\cosh\alpha \cosh\phi -\frac{1}{2}\sinh\alpha \sinh\phi \sin \theta ]
^\frac{1}{2}}, \\
B(\phi ,\alpha ,\theta )&=&
\frac{
\sinh\frac{\alpha }{2}\sinh\frac{\phi}{2}
}{
[\frac{1}{2}+\frac{1}{2}\cosh\alpha \cosh\phi -\frac{1}{2}\sinh\alpha \sinh\phi \sin \theta ]
^\frac{1}{2}
}  \cos \theta .
\end{eqnarray}
Hence, 
\begin{eqnarray}
U(\Lambda) |\varphi _p(\theta )\rangle =\frac{1}{\sqrt{2}} \left[
	a_{\Lambda p}(\theta )
	\left(
	|\Lambda p(\theta ),\uparrow \rangle |\Lambda p(\theta +\pi),\downarrow \rangle 
	-|\Lambda p(\theta ),\downarrow \rangle |\Lambda p(\theta +\pi),\uparrow \rangle 
	\right) 
	\right. \nonumber \\
	+
	\left.
	b_{\Lambda p}(\theta )
	\left(
	|\Lambda p(\theta ),\uparrow \rangle |\Lambda p(\theta +\pi),\uparrow \rangle 
	+|\Lambda p(\theta ),\downarrow \rangle |\Lambda p(\theta +\pi),\downarrow \rangle 
	\right)
	\right], \nonumber \\ \label{spin_singlet_at_laboratory}
\end{eqnarray}
where 
\begin{eqnarray}
a_{\Lambda p}(\theta )
&=&
	A(\phi ,\alpha ,\theta )A(\phi ,\alpha ,\theta+\pi)+B(\phi ,\alpha ,\theta )B(\phi ,\alpha ,\theta+\pi )  \nonumber 
	\\
&=&
	\frac{\cosh\alpha +\cosh\phi }
	{[(1+\cosh\alpha \cosh\phi )^2-\sinh^2\alpha \sinh^2\phi \sin^2\theta ]^\frac{1}{2}}
	\label{a_lambda_p1}, \\
b_{\Lambda p}(\theta )
&=&
	B(\phi ,\alpha ,\theta )A(\phi ,\alpha ,\theta+\pi)-A(\phi ,\alpha ,\theta )B(\phi ,\alpha ,\theta+\pi ) \nonumber \\
&=&
	\frac{\sinh\alpha \sinh\phi }
	{[(1+\cosh\alpha \cosh\phi )^2-\sinh^2\alpha \sinh^2\phi \sin^2\theta ]^\frac{1}{2}} \cos\theta 
	\label{b_lambda_p1},
\end{eqnarray}
and these satisfy $|a_{\Lambda p}(\theta )|^2+|b_{\Lambda p}(\theta )|^2=1$. Here, we have introduced a new notation such as $a_{\Lambda p}(\theta )$ and $b_{\Lambda p}(\theta )$ instead of $a(\phi ,\alpha ,\theta )$ and $b(\phi ,\alpha ,\theta )$.
We notice also that Eq. (\ref{a_lambda_p1}) and Eq. (\ref{b_lambda_p1}) can be rewritten in terms of $v$ and $\alpha $ as 
\begin{eqnarray}
a_{\Lambda p}(\theta )	&=&
	\frac{1+\sqrt{1-v^2}\cosh\alpha }
	{[(\sqrt{1-v^2}+\cosh\alpha )^2-v^2 \sinh^2\alpha \sin^2\theta ]^\frac{1}{2}}, 
	\label{a_lambda_p2} \\
b_{\Lambda p}(\theta )
&=&
	\frac{v \sinh\alpha }
	{[(\sqrt{1-v^2}+\cosh\alpha )^2-v^2 \sinh^2\alpha \sin^2\theta ]^\frac{1}{2}} \cos\theta . 
	\label{b_lambda_p2}
\end{eqnarray}
We can also describe $a_{\Lambda p}(\theta )$ and $b_{\Lambda p}(\theta )$ in terms of $V$ and $\phi$, by replacement of $v\rightarrow V$ and $\alpha \rightarrow \phi $. Eq. (\ref{a_lambda_p1}) and Eq. (\ref{b_lambda_p1}) are found to be symmetric for an exchange of $\alpha $ and $\phi$. 
From Eq. (\ref{spin_singlet_at_laboratory}), we can see that the spin-singlet state in the center of mass frame becomes the superposition state of spin-singlet state and $S_z=\pm 1$ spin-triplet states in the laboratory frame. 
In the non-relativistic limit $v\rightarrow 0$, Eq. (\ref{a_lambda_p2}) and Eq. (\ref{b_lambda_p2}) become
\begin{eqnarray}
a_{\Lambda p}(\theta )	&=&	1-\frac{1}{2}v^2 \cos^2 \theta \tanh^2 \frac{\alpha }{2} +O(v^3), \\
b_{\Lambda p}(\theta )	&=&	v \cos \theta \tanh \frac{\alpha }{2}+O(v^3).
\end{eqnarray}
and we can see how the fraction of the spin-triplet state increases as a relativistic effect.
But, for $\theta =\frac{\pi}{2}$, since $a_{\Lambda p}(\frac{\pi}{2} )=1$ and $b_{\Lambda p}(\frac{\pi}{2} )=0$, the spin state in the laboratory frame remains in the pure spin-singlet state as 
\begin{eqnarray}
U(\Lambda) |\varphi _p(\frac{\pi }{2} )\rangle =\frac{1}{\sqrt{2}}
	(
	|\Lambda p(\frac{\pi }{2} ),\uparrow \rangle |\Lambda p(\frac{\pi }{2} +\pi),\downarrow \rangle 
	-|\Lambda p(\frac{\pi }{2} ),\downarrow \rangle |\Lambda p(\frac{\pi }{2} +\pi),\uparrow \rangle 
	). \nonumber \\ \label{spin_singlet_in_laboratory_at_pi/2}
\end{eqnarray}

Next we consider the super-relativistic limit where the ejection velocity of the particles is $v\rightarrow 1$ and the boost velocity is $V\rightarrow 1$. Since $a_{\Lambda p}(\theta )\rightarrow 0$ and $b_{\Lambda p}(\theta )\rightarrow 1$ ($\theta < \frac{\pi }{2}$) or $b_{\Lambda p}(\theta )\rightarrow -1$ ($\theta > \frac{\pi }{2}$) in this limit, the state in the laboratory frame converges to 
\begin{eqnarray}
U(\Lambda) |\varphi _p(\theta )\rangle \rightarrow \beta \frac{1}{\sqrt{2}}
	\left(
	|\Lambda p(\theta ),\uparrow \rangle |\Lambda p(\theta +\pi),\uparrow \rangle 
	+|\Lambda p(\theta ),\downarrow \rangle |\Lambda p(\theta +\pi),\downarrow \rangle 
	\right), \nonumber \\
	\label{spin_singlet_in_laboratory_at_super_rela}
\end{eqnarray}
where $\beta = +1$ for $\theta < \frac{\pi }{2}$ and $\beta =-1$ for $\theta > \frac{\pi }{2}$. This is the superposition state of $S_z=\pm 1$ spin-triplet states. 

In the following of this paper, we call this superposition state of Eq. (\ref{spin_singlet_in_laboratory_at_super_rela}) as the "triplet-state" simply. 
The triplet-state is one of the Bell states as well as the singlet-state.

\section{The von Neumann entropy of EPR spin correlation in the laboratory frame}\label{The von Neumann entropy of spin correlation in the laboratory frame}

Using the unitary operator $U(\Lambda )$ and the state vector in the center of mass frame such as
\begin{eqnarray}
|\psi _p(\theta_1,\theta_2,\cdots, \theta_i ,\cdots )\rangle =
\sum _i c_i|\varphi _p(\theta_i)\rangle \:\text{with}\: \sum _i|c_i|^2=1, \label{state_vector_mixed_with_two_theta_in_com}
\end{eqnarray}
we can represent the density matrix in the laboratory frame as follows:
\begin{eqnarray}
\rho _{\Lambda p}(\theta_1,\theta_2,\cdots, \theta_i ,\cdots )= 
	U(\Lambda )|\psi _p(\theta_1,\theta_2,\cdots, \theta_i ,\cdots )\rangle
	\langle \psi _p(\theta_1,\theta_2,\cdots, \theta_i ,\cdots )|U^\dag (\Lambda ). \nonumber \\
\label{the density matrix in the laboratory frame}
\end{eqnarray}
Taking the trace over 4-momentum, the reduced density matrix is obtained as 
\begin{eqnarray}
\rho' _{\Lambda p}(\theta_1,\theta_2,\cdots, \theta_i ,\cdots )&=&
	\sum _i p_i M_{\Lambda p}(\theta_i),
\label{reduced_d_matrix_in_laboratory} \\
M_{\Lambda p}(\theta _i)&=&\text{Tr}_{\text{4m}}
[U(\Lambda )|\varphi _p(\theta_i)\rangle \langle \varphi _p(\theta_i)|U^\dag(\Lambda )],
\label{the reduced density matrix of one pair state}
\end{eqnarray}
where $p_i$ denotes $|c_i|^2$ and $\text{Tr}_{\text{4m}}[\cdots]$ expresses the trace over 4-momentum. The value of this trace depends both on $\Lambda p$ and $p$. Because, since $a_{\Lambda p}(\theta _i)$ and $b_{\Lambda p}(\theta _i)$ in $U(\Lambda )|\varphi _p(\theta_i)\rangle \langle \varphi _p(\theta_i)|U^\dag(\Lambda )$ depend on the 4-momentums, the operation of trace implies that $\Lambda p$ and $p$ are picked up in any 4-momentums; the reduced density matrix Eq. (\ref{the reduced density matrix of one pair state}) is parameterized with the picked-up $\Lambda p$ and $p$. 

From the reduced density matrix Eq. (\ref{reduced_d_matrix_in_laboratory}), we calculate the von Neumann entropy as follows: 
\begin{eqnarray}
S(\phi, \alpha, \theta_1,\theta_2,\cdots, \theta_i ,\cdots )
=
-\text{Tr}[\rho' _{\Lambda p}(\theta_1,\theta_2,\cdots, \theta_i ,\cdots )\log \rho' _{\Lambda p}(\theta_1,\theta_2,\cdots, \theta_i ,\cdots )].
\nonumber \\
\end{eqnarray}
Here, "taking the trace" means to ignore the coupling of the 4-momentum and the spin. In the center of mass frame, the von Neumann entropy is always 0 since the reduced density matrix is the pure state. 
In this paper, we are discussing the entropy for the spin states classified by $ \{|\uparrow \uparrow \rangle $, $|\uparrow \downarrow \rangle $, $|\downarrow \uparrow \rangle $, $|\downarrow \downarrow \rangle \}$. 

In the following, we consider the simplest case of two pair states and we set $\theta _1=0$ and $\theta _2=\theta $. 
Then, using
\begin{eqnarray}
|\psi _p(0, \theta)\rangle &=&
	c_1|\varphi _p(0)\rangle +c_2|\varphi _p(\theta )\rangle 
	\label{superposed spin-singlet in the frame A with 0 and theta} \\
\text{and}~~\rho' _{\Lambda p}(0,\theta)&=&
	p_1 M_{\Lambda p}(0)+p_2 M_{\Lambda p}(\theta), 
\label{dash_reduced_d_matrix_in_laboratory}
\end{eqnarray}
the von Neumann entropy is calculated as
 \begin{eqnarray}
S(\phi ,\alpha , 0, \theta )=-\text{Tr}[\rho' _{\Lambda p}(0,\theta)\log \rho' _{\Lambda p}(0,\theta)].
\label{the von Neumann entropy of distinguishable state 0 and theta}
\end{eqnarray}

\begin{figure}[htbp]
	\parbox{\halftext}{
		\centerline{\includegraphics[width=6.6cm]{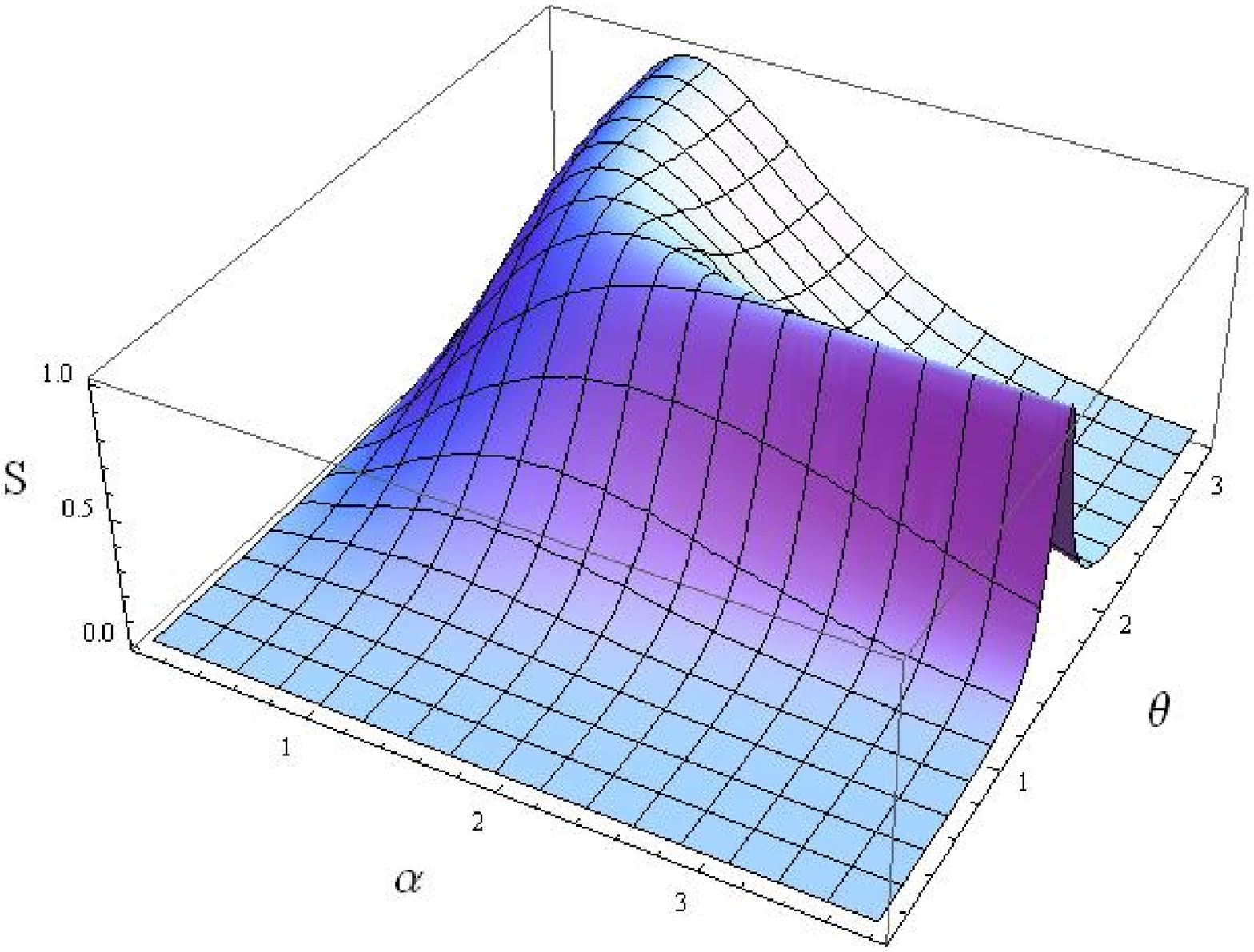}}
	\caption{The von Neumann entropy of the state $\rho '_{\Lambda p}(0,\theta)$ where $p_1=p_2=\frac{1}{2}$, $\phi=\tanh^{-1}0.999$; $0.1 \leq  \alpha \leq \tanh^{-1}0.999$, $0.1 \leq \theta \leq \pi$. }
	\label{neumann 1to1}}
	\hfill
	\parbox{\halftext}{
		\centerline{\includegraphics[width=6.6cm]{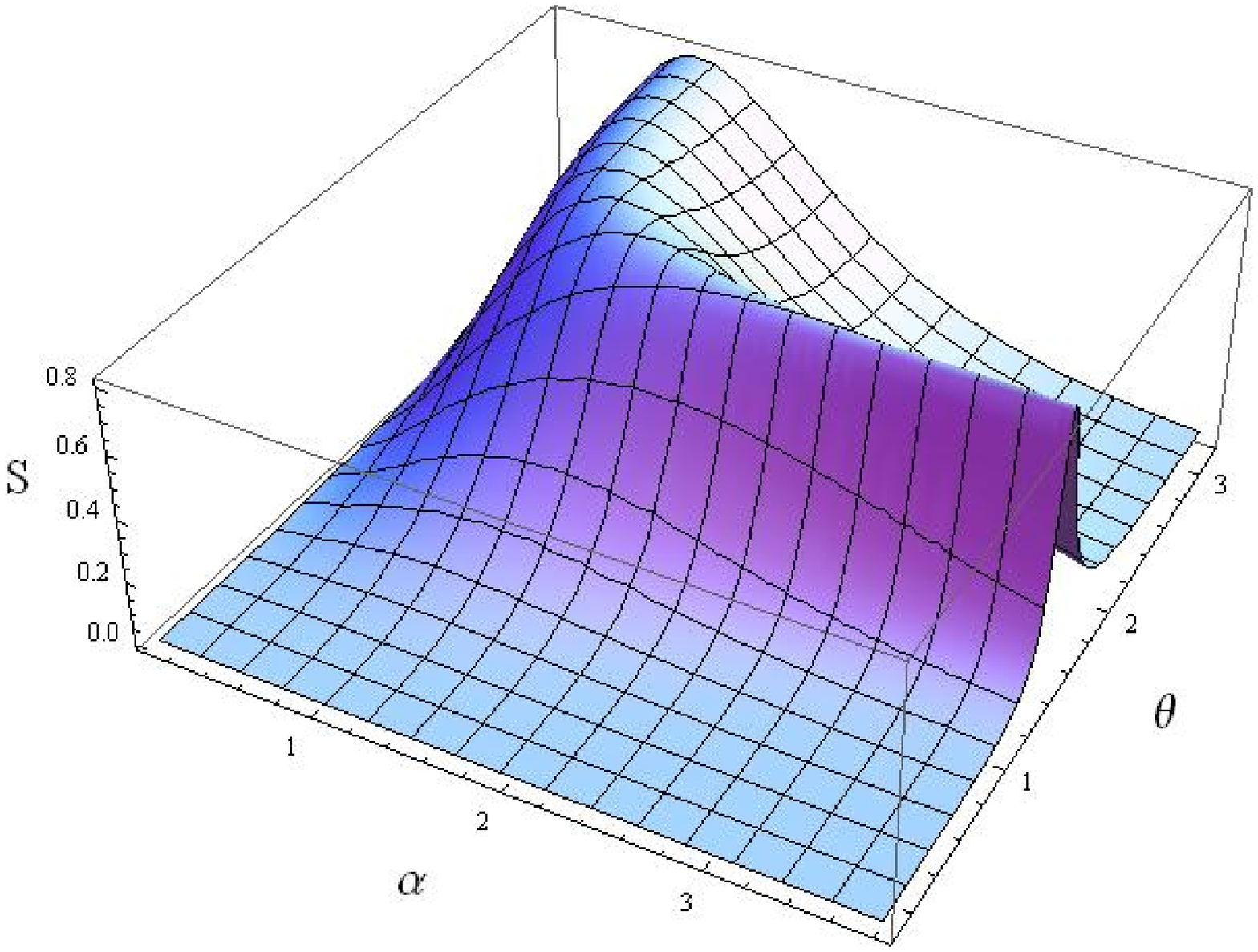}}
	\caption{The von Neumann entropy of the state $\rho '_{\Lambda p}(0,\theta)$ where $p_1=\frac{1}{4}$, $p_2=\frac{3}{4}$, $\phi=\tanh^{-1}0.999$; $0.1 \leq  \alpha \leq \tanh^{-1}0.999$, $0.1 \leq \theta \leq \pi$. }
	\label{neumann 1to3}}
\end{figure}

We have computed the above equation for $S$ numerically by the Mathematica and the results are shown in the figures: Fig. \ref{neumann 1to1} for the case of $p_1=p_2=\frac{1}{2}$ and Fig. \ref{neumann 1to3} for the case of $p_1=\frac{1}{4}$, $p_2=\frac{3}{4}$.

From Figs. \ref{neumann 1to1} and \ref{neumann 1to3}, we can see that $S$ tends to 0 or a finite value in the super-relativistic limit. For the general cases of $\theta \ne \frac{\pi }{2}$, $S$ tends to $0$. We can explain this phenomenon from Eq. (\ref{spin_singlet_in_laboratory_at_super_rela}); in the laboratory frame, the state vector converges to the pure state as the "triplet-state" and $S$ takes the minimum value of $0$. This conforms with the fact that a massless particle's spin direction is parallel to the moving direction of itself. 

At $\theta =\frac{\pi }{2}$, on the other hand, $S$ converges to a finite value in the super-relativistic limit, where the singlet-state $|\varphi _p(0) \rangle $ becomes the triplet-state in the laboratory frame but $|\varphi _p(\frac{\pi }{2})\rangle $ remains the singlet-state irrespective of $\alpha $ and $\phi$.

\subsection{The maximum value of the von Neumann entropy}

In Figs. \ref{neumann 1to1} and \ref{neumann 1to3}, one feature of the variation of $S$ is not symmetric with respect to $\frac{\pi }{2}$. The value of $S$ at $\theta=\frac{\pi}{2}$ converges to the maximum value $S_{\text{max}}$ in the super-relativistic limit. Another feature is that, in the region of $\theta >\frac{\pi}{2}$, $S$ increases with $\alpha$, takes $S_{\text{max}}$ at some $\alpha$ and decreases again in the super-relativistic limit. The points of $S_{\text{max}}$ depend on $\phi$, $\alpha $ and $\theta$.
Since the von Neumann entropy is a measure of the purity of the state, the maximum value expresses how much the purity is degraded. 

We can represent the density matrix Eq. (\ref{dash_reduced_d_matrix_in_laboratory}) using the basis set composed of the singlet-state and the triplet-state as follows: 
\begin{eqnarray}
\rho '_{\Lambda p}(0,\theta )
=
\left(
\begin{array}{cc}
 	p_1 |a_{\Lambda p}(0)|^2+ p_2 |a_{\Lambda p}(\theta )|^2& 	p_1 a_{\Lambda p}(0)b_{\Lambda p}^*(0)+ p_2 a_{\Lambda p}(\theta )b_{\Lambda p}^*(\theta )\\
 	p_1 a_{\Lambda p}^*(0)b_{\Lambda p}(0)+ p_2 a_{\Lambda p}^*(\theta )b_{\Lambda p}(\theta )& p_1 |b_{\Lambda p}(0)|^2+ p_2 |b_{\Lambda p}(\theta )|^2
\end{array}
\right), \nonumber \\
\label{the reduced density matrix 0 and theta}
\end{eqnarray}
where $a_{\Lambda p}(\theta )$ and $b_{\Lambda p}(\theta )$ are given in Eq. (\ref{a_lambda_p1}) and Eq. (\ref{b_lambda_p1}). 

In the region of $\theta >\frac{\pi }{2}$, $S$ takes the maximum value at some points of $\alpha$ and $\phi$ as mentioned before. We can explain this behavior in the following: From Eq. (\ref{the reduced density matrix 0 and theta}), it is found that 
\begin{eqnarray}
\text{det}[\rho '_{\Lambda p}(0, \theta)]=
	p_1 p_2
	|a_{\Lambda p}(\theta)|^2 |b_{\Lambda p}(0)|^2 (\cos \theta-1)^2.
	\label{det rho}
\end{eqnarray}
We notice first that $0 \leq \text{det}[\rho '_{\Lambda p}(0,\theta )] \leq p_1 p_2$; and, since the eigenvalues $\lambda _1$ and $\lambda _2$ of $\rho '_{\Lambda p}(0,\theta )$ are given as
\begin{eqnarray}
\lambda_1 = \frac{1}{2}+\frac{\sqrt{1-4 \text{det}[\rho '_{\Lambda p}(0,\theta )]}}{2}, \label{lambda1} \\
\lambda_2 = \frac{1}{2}-\frac{\sqrt{1-4 \text{det}[\rho '_{\Lambda p}(0,\theta )]}}{2}, \label{lambda2}
\end{eqnarray}
it is found that $S$ takes the maximum value when $\text{det}[\rho '_{\Lambda p}(0,\theta )]=p_1 p_2$. 

The requirement of $\text{det}[\rho '_{\Lambda p}(0,\theta )]=p_1 p_2$ induces the condition
\begin{eqnarray}
	|a_{\Lambda p}(\theta)|^2 |b_{\Lambda p}(0)|^2 (\cos \theta-1)^2
	=1,
\end{eqnarray}
which is satisfied under the following relation among $\alpha$, $\phi$ and $\theta$
\begin{eqnarray}
\left(
\frac{\cosh \alpha + \cosh \phi }{\sinh \alpha \sinh \phi}
\right)^2
=
-\cos \theta ~~~\text{for}~~ \theta > \frac{\pi}{2}. 
\label{alpha_phi_relation}
\end{eqnarray}
We notice that this relation is symmetric for $\alpha$ and $\phi$.
When $\text{det}[\rho '_{\Lambda p}(0, \theta )]=p_1 p_2$, $\lambda_1=p_1$ and $\lambda _2=p_2$ from Eq. (\ref{lambda1}) and Eq. (\ref{lambda2}). Thus, the maximum value is
\begin{eqnarray}
S_{\text{max}}(0, \theta)=-p_1 \log p_1 - p_2 \log p_2 ~~~\text{for}~~\theta > \frac{\pi }{2}. \label{the extreme value of distinguishable state}
\end{eqnarray}

For $\theta>\frac{\pi }{2}$, $\alpha $ at the $S_{\text{max}}$ is given by the following equation: 
\begin{eqnarray}
\cosh \alpha =\frac{-\cosh\phi-\sinh^2\phi\sqrt{-\cos \theta(1-\cos \theta)}}{1+\cos \theta \sinh^2\phi}.
\label{phi->alpha}
\end{eqnarray}
In the limit of $\phi \rightarrow \infty $ ($v \rightarrow 1$), the rapidity of the relative velocity tends to
\begin{eqnarray}
\cosh \alpha \rightarrow \sqrt{1-\frac{1}{\cos \theta }} ~~~\text{for}~~\theta > \frac{\pi }{2}, 
\end{eqnarray}
and, as $\tanh \alpha = \sqrt{\frac{1}{1-\cos \theta }}$, the $\alpha $ at the maximum value lies in the range of 
\begin{eqnarray}
\tanh^{-1}\sqrt{\frac{1}{1-\cos \theta }}<\alpha <\infty  ~~~\text{for}~~\theta > \frac{\pi }{2}. 
\end{eqnarray}
Then, the $\alpha $ decreases with $\theta $ up to $\pi $ and, in $\theta = \pi$, $S$ has the maximum value when the relative velocity takes the minimum value. 
In the same manner, we can also describe $\cosh\phi$ using $\alpha $ by the replacement of $\phi \leftrightarrow \alpha$ in Eq. (\ref{phi->alpha}).

We can also rewrite Eq. (\ref{alpha_phi_relation}) as 
\begin{eqnarray}
\frac{a_{\Lambda p}(0)}{b_{\Lambda p}(0)} \frac{a_{\Lambda p}(\theta)}{b_{\Lambda p}(\theta)}=-1,
\end{eqnarray}
and, as $a_{\Lambda p}(0)=-b_{\Lambda p}(\theta)$ and $b_{\Lambda p}(0)=a_{\Lambda p}(\theta)$, 
\begin{eqnarray}
M_{\Lambda p}(0) M_{\Lambda p}(\theta)=0. 
\end{eqnarray}
Therefore, when the von Neumann entropy becomes Eq. (\ref{the extreme value of distinguishable state}), the state for $M_{\Lambda p}(0)$ is orthogonal to the state for $M_{\Lambda p}(\theta)$; in other words, the reduced density matrix $\rho '_{\Lambda p}(0, \theta)$ is diagonalized with the state for $M_{\Lambda p}(0)$ and the state for $M_{\Lambda p}(\theta)$. 

\section{The von Neumann entropy and the Shannon entropy}\label{The von Neumann entropy and the Shannon entropy}

We can calculate the Shannon entropy in the laboratory frame for our extended model. The Shannon entropy corresponds to the von Neumann entropy of the measured data. Using the following probabilities, which are obtained immediately from the reduced density matrix $\rho '_{\Lambda p}(0,\theta )$ represented with the basis set $\{|\uparrow \uparrow \rangle , |\uparrow \downarrow \rangle , |\downarrow \uparrow \rangle , |\downarrow \downarrow \rangle \}$,
\begin{eqnarray}
P(|\uparrow \downarrow \rangle )=P(|\downarrow \uparrow \rangle )&=&\frac{p_1 |a_{\Lambda p}(0)|^2+ p_2 |a_{\Lambda p}(\theta )|^2}{2}, \label{the probabilitys of updown and downup}\\
P(|\uparrow \uparrow \rangle )=P(|\downarrow \downarrow \rangle )&=&\frac{p_1 |b_{\Lambda p}(0)|^2+ p_2 |b_{\Lambda p}(\theta )|^2}{2}, \label{the probabilitys of upup and downdown}
\end{eqnarray}
the Shannon entropy is calculated as 
\begin{eqnarray}
S_{\text{Sh}}(0, \theta)=- \left( p_1 |a_{\Lambda p}(0)|^2 + p_2 |a_{\Lambda p}(\theta )|^2 \right) \log \left(p_1 |a_{\Lambda p}(0)|^2 + p_2 |a_{\Lambda p}(\theta )|^2 \right) \nonumber \\
-\left(p_1 |b_{\Lambda p}(0)|^2+p_2 |b_{\Lambda p}(\theta )|^2 \right)\log \left(p_1 |b_{\Lambda p}(0)|^2 + p_2 |b_{\Lambda p}(\theta )|^2 \right) +1 .
\label{the Shannon entropy}
\end{eqnarray}
This is symmetric with respect to $\frac{\pi }{2}$, because the $\theta$ is contained in only the form of $\sin ^2 \theta$ and $\cos ^2 \theta$ in Eq. (\ref{the Shannon entropy}). 

\begin{figure}[htbp]
	\parbox{\halftext}{
		\centerline{\includegraphics[width=6.6cm]{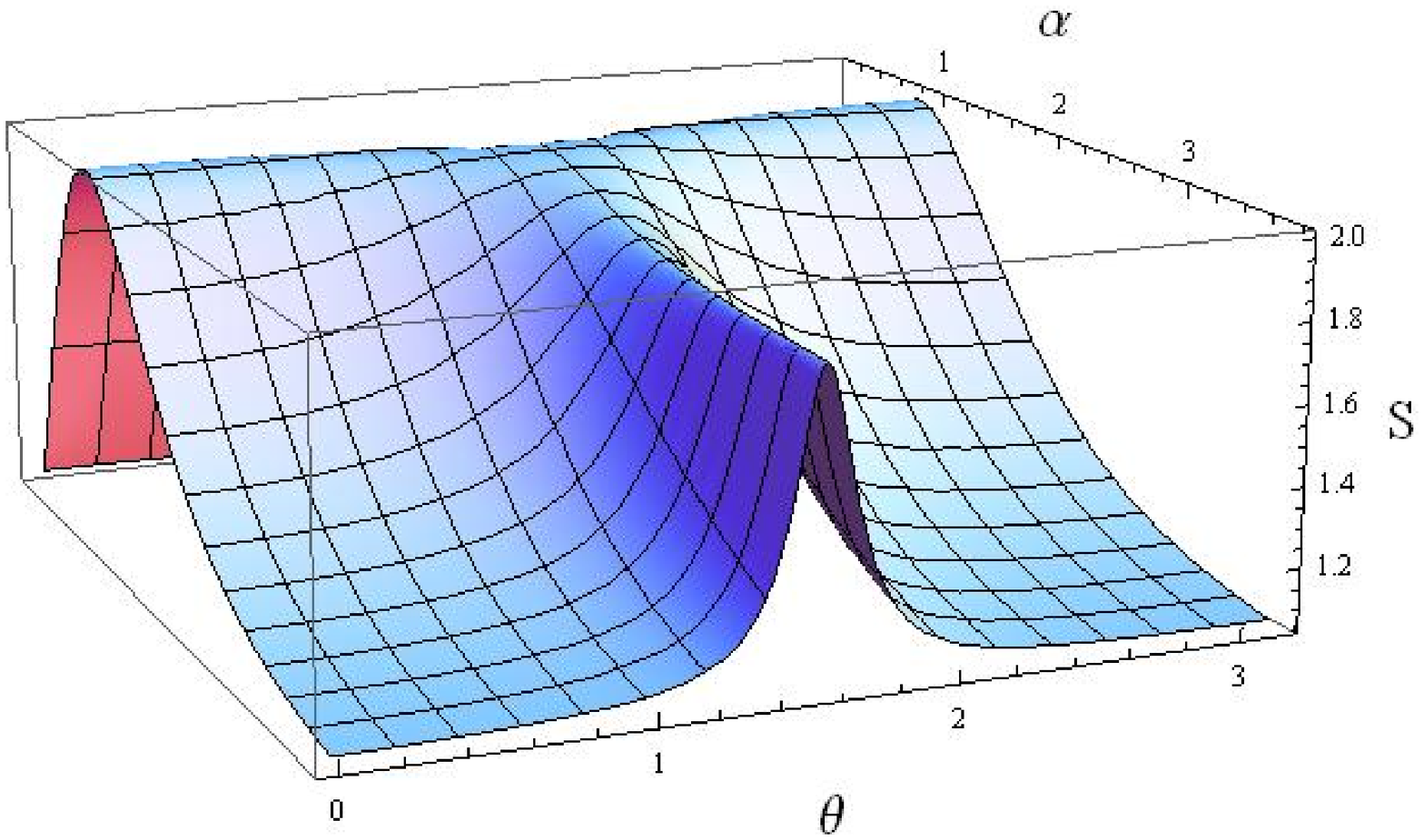}}
	\caption{The Shannon entropy of the spin correlation where $p_1=\frac{3}{4}$, $p_2=\frac{1}{4}$, $\phi=\tanh^{-1}0.999$; $0.1 \leq  \alpha \leq \tanh^{-1}0.999$, $0.1 \leq \theta \leq \pi$. }
	\label{shannon 3to1}}
	\hfill
	\parbox{\halftext}{
		\centerline{\includegraphics[width=6.6cm]{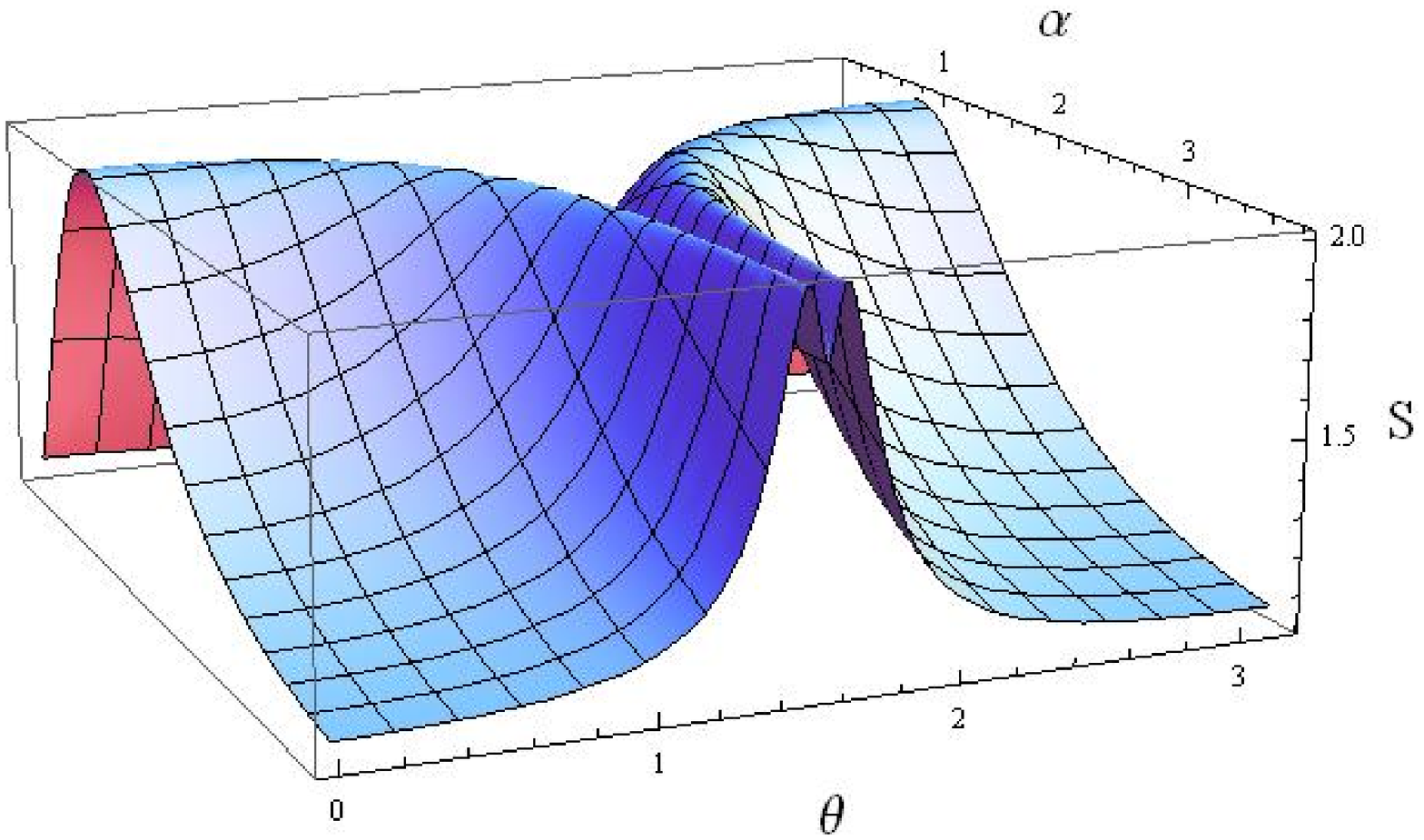}}
	\caption{The Shannon entropy of the spin correlation where $p_1=\frac{1}{4}$, $p_2=\frac{3}{4}$, $\phi=\tanh^{-1}0.999$; $0.1 \leq  \alpha \leq \tanh^{-1}0.999$, $0.1 \leq \theta \leq \pi$. }
	\label{shannon 1to3}}
\end{figure}

\begin{figure}[htbp]
	\parbox{\halftext}{
		\centerline{\includegraphics[width=6.6cm]{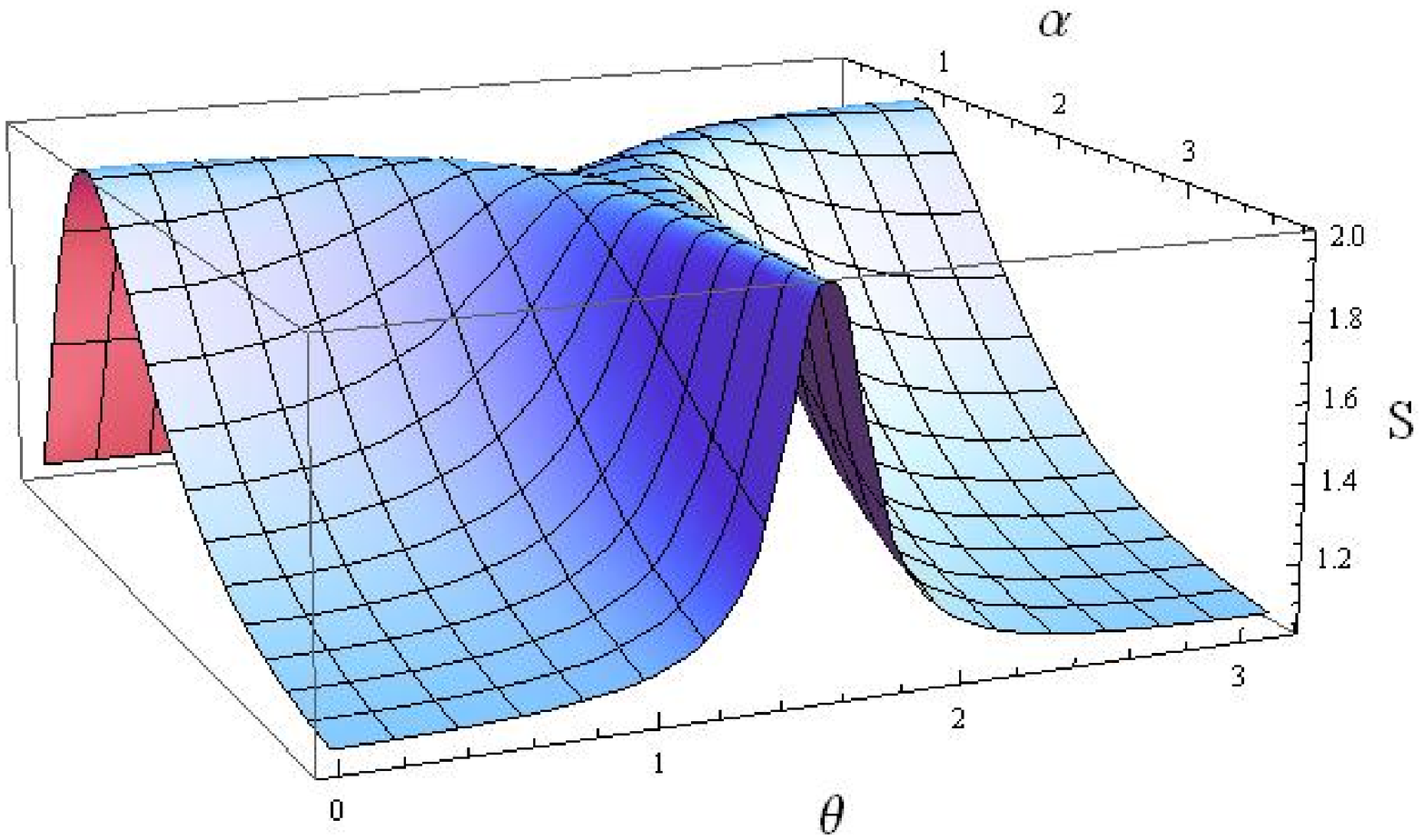}}
	\caption{The Shannon entropy of the spin correlation where $p_1=p_2=\frac{1}{2}$, $\phi=\tanh^{-1}0.999$; $0.1 \leq  \alpha \leq \tanh^{-1}0.999$, $0.1 \leq \theta \leq \pi$. }
	\label{shannon 1to1}}
	\hfill
	\parbox{\halftext}{
		\centerline{\includegraphics[width=6.6cm]{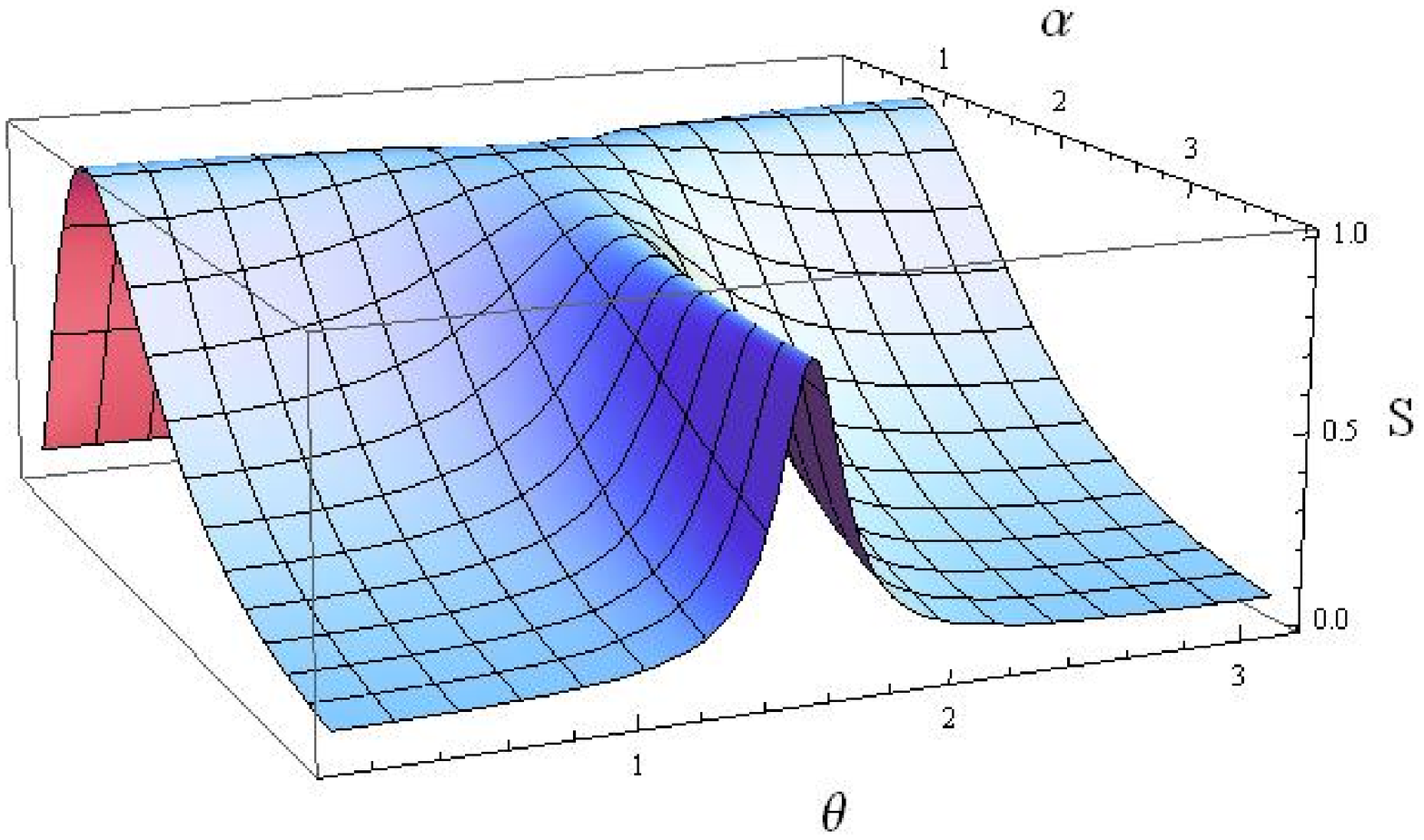}}
	\caption{The von Neumann entropy of the indistinguishable state $\rho '_{\Lambda p}(0,\theta,\pi,\pi+\theta)$ where $p_1=\frac{3}{4}$, $p_2=\frac{1}{4}$, $\phi=\tanh^{-1}0.999$; $0.1 \leq  \alpha \leq \tanh^{-1}0.999$, $0.1 \leq \theta \leq \pi$. }
	\label{neumann-id 3to1}}
\end{figure}

We have shown the results in Figs. \ref{shannon 3to1} $\sim $ \ref{shannon 1to1} for the cases of ($p_1=\frac{3}{4}$, $p_2=\frac{1}{4}$ ), ($p_1=\frac{1}{4}$, $p_2=\frac{3}{4}$ ) and ($p_1=p_2=\frac{1}{2}$ ).

\subsection{Indistinguishable state}

Suppose that we label the particle moving "to the right" as Particle A and one "to the left" as Particle B for $\theta = 0$. However, for $\theta > \frac{\pi }{2}$, Particle A is moving "to the left" and Particle B does "to the right". Therefore, if the state at $\theta > \frac{\pi }{2}$ was not reduced to the state at $0<\theta <\frac{\pi }{2}$ as for the case of Eq. (\ref{superposed spin-singlet in the frame A with 0 and theta}), it implicitly implies that Particle A and Particle B are distinguishable. Thus, the von Neumann entropy gives asymmetric value with respect to $\theta =\frac{\pi }{2}$ different from the Shannon entropy. 

\begin{figure}[htbp]
	\parbox{\halftext}{
		\centerline{\includegraphics[width=6.6cm]{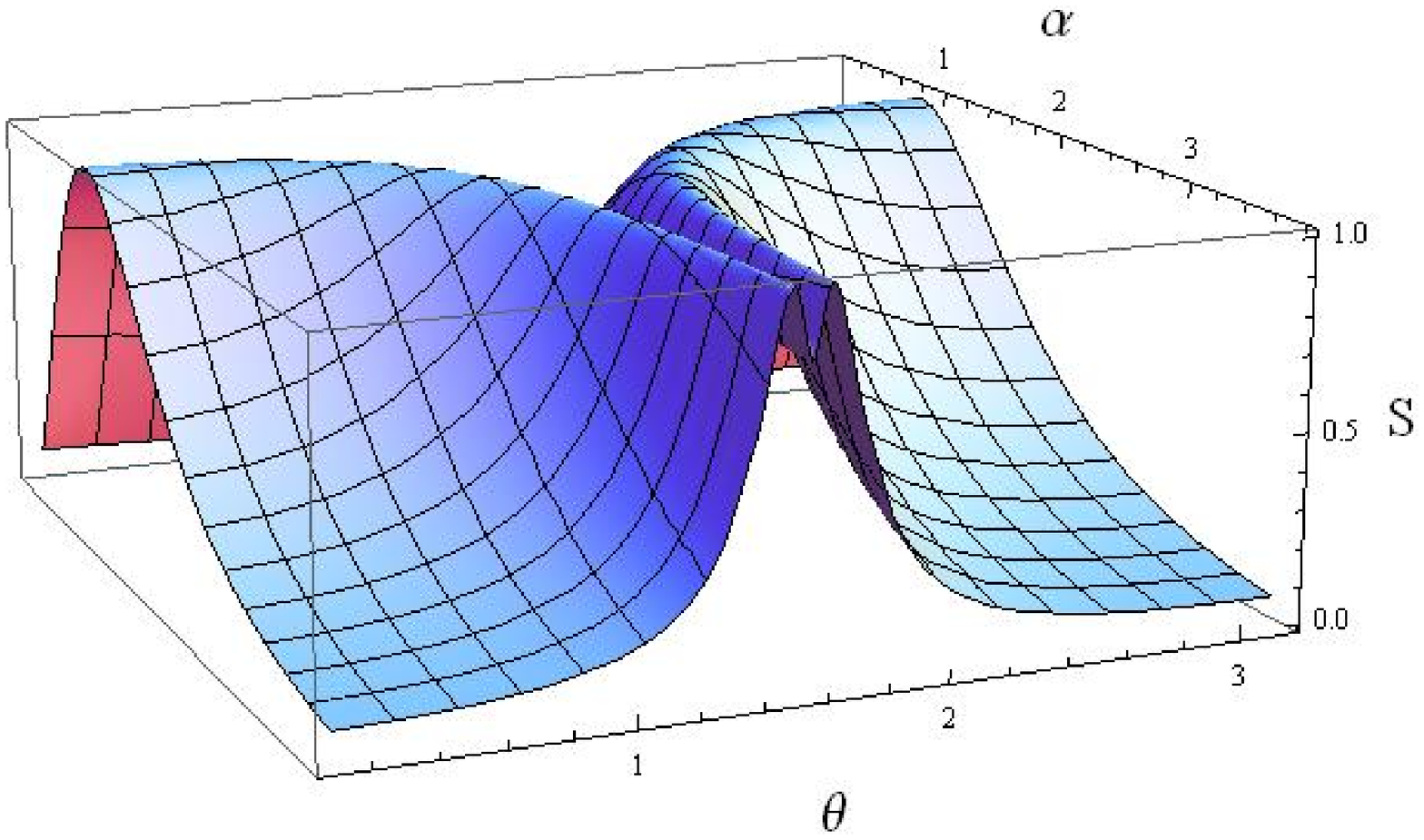}}
	\caption{The von Neumann entropy of the indistinguishable state $\rho '_{\Lambda p}(0,\theta,\pi,\pi+\theta)$ where $p_1=\frac{1}{4}$, $p_2=\frac{3}{4}$, $\phi=\tanh^{-1}0.999$; $0.1 \leq  \alpha \leq \tanh^{-1}0.999$, $0.1 \leq \theta \leq \pi$. }
	\label{neumann-id 1to3}}
	\hfill
	\parbox{\halftext}{
		\centerline{\includegraphics[width=6.6cm]{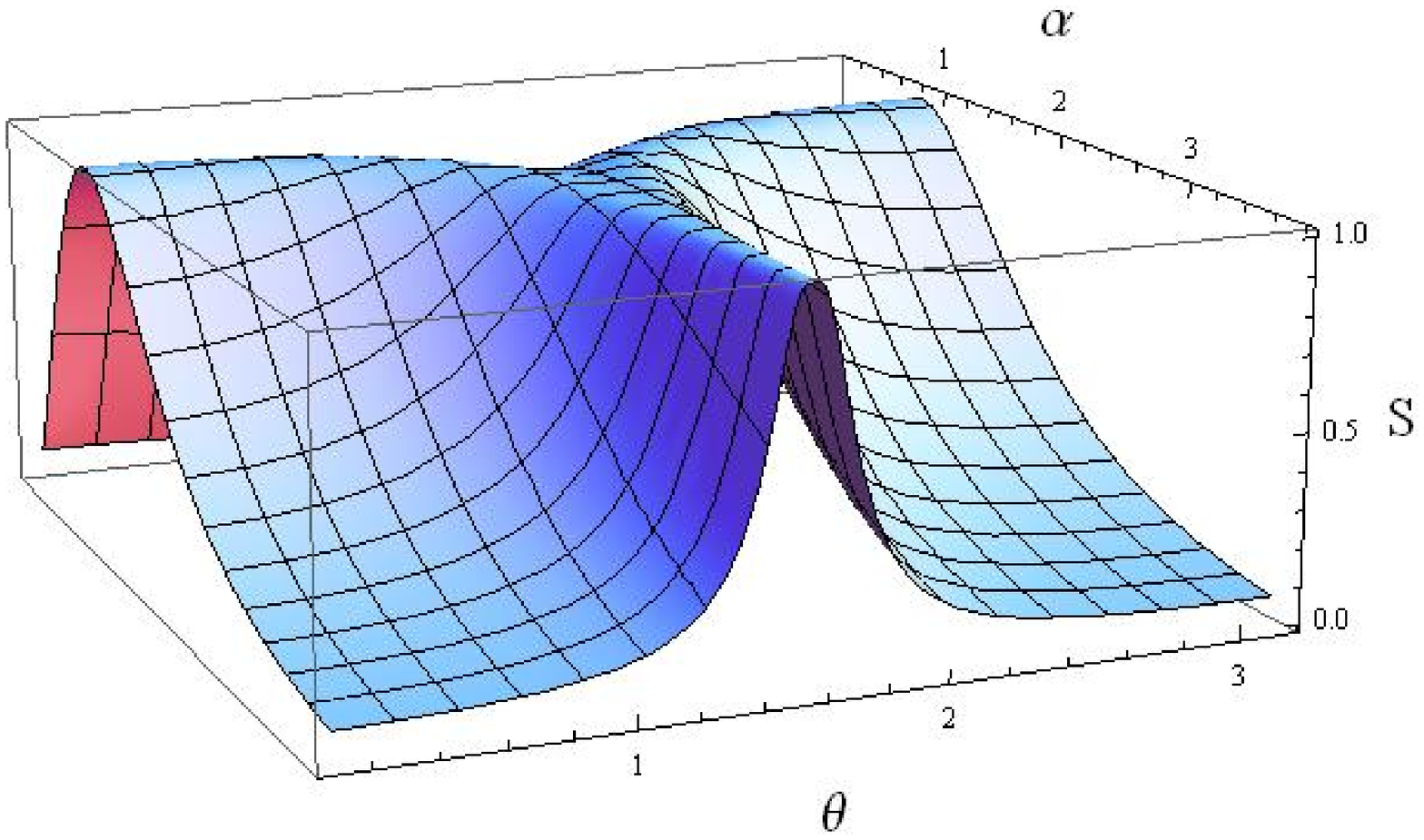}}
	\caption{The von Neumann entropy of the indistinguishable state $\rho '_{\Lambda p}(0,\theta,\pi,\pi+\theta)$ where $p_1=p_2=\frac{1}{2}$, $\phi=\tanh^{-1}0.999$; $0.1 \leq  \alpha \leq \tanh^{-1}0.999$, $0.1 \leq \theta \leq \pi$. }
	\label{neumann-id 1to1}}
\end{figure}

If Particle A and Particle B are indistinguishable, we should prepare that the state vector is antisymmetric in spatial coordinate, instead of Eq. (\ref{superposed spin-singlet in the frame A with 0 and theta}). We call the antisymmetric state the indistinguishable state and the original state the distinguishable state. The indistinguishable state for the superposition state of two pair states can be written by introducing the four pair states in Eq. (\ref{state_vector_mixed_with_two_theta_in_com}), imposing the antisymmetric condition: $c_1=-c_3$, $c_2=-c_4$ and $\theta_1=0$, $\theta_2=\theta $, $\theta_3=\pi $, $\theta_4=\pi +\theta$. Note that $\sum |c_i|^2=2$ because we have set that $c_1$ and $c_2$ are equal in Eq. (\ref{superposed spin-singlet in the frame A with 0 and theta}). Thus the state vector is given as
\begin{eqnarray}
|\psi_{p}(0,\theta ,\pi, \pi+\theta) \rangle =\frac{1}{\sqrt{2}}\left[ \left(c_1 |\varphi_{p}(0)\rangle +c_2 |\varphi_{p}(\theta )\rangle \right)
- \left( c_1 |\varphi_{p}(\pi )\rangle +c_2 |\varphi_{p}(\pi +\theta )\rangle \right)\right]. \nonumber \\
\label{no-distinguished particles state}
\end{eqnarray}
We have calculated the von Neumann entropy in the laboratory frame from this state vector and the results have been shown in Figs. \ref{neumann-id 3to1}$\sim $\ref{neumann-id 1to1}. It is found that the calculated von Neumann entropy coincides with the Shannon entropy, except for the difference of absolute value by 1.

The reduced density matrix $\rho '_{\Lambda p}(0, \theta,\pi,\pi+\theta)$ from Eq. (\ref{no-distinguished particles state}) can be diagonalized by taking the basis set composed of the singlet-state and the triplet-state:
\begin{eqnarray}
\rho '_{\Lambda p}(0,\theta ,\pi, \pi+\theta)
&=&
\left( 
\begin{array}{cc}
 	p_1 |a_{\Lambda p}(0)|^2+ p_2 |a_{\Lambda p}(\theta )|^2& 	0 \\
 	0&	 p_1 |b_{\Lambda p}(0)|^2+ p_2 |b_{\Lambda p}(\theta )|^2
\end{array}
\right). \nonumber \\
\label{the reduced density matrix of indistinguishable state}
\end{eqnarray}
Therefore, we can regard the state to be the mixed state of the triplet-state and the singlet-state. And then the von Neumann entropy becomes as follows: 
\begin{eqnarray}
S_{\text{vN}} (0, \theta) &=& - \left( p_1 |a_{\Lambda p}(0)|^2+p_2 |a_{\Lambda p}(\theta )|^2 \right) \log \left(p_1 |a_{\Lambda p}(0)|^2 + p_2 |a_{\Lambda p}(\theta )|^2 \right) \nonumber \\
&&
-\left(p_1|b_{\Lambda p}(0)|^2+p_2 |b_{\Lambda p}(\theta )|^2 \right)\log \left(p_1 |b_{\Lambda p}(0)|^2 + p_2 |b_{\Lambda p}(\theta )|^2 \right) \label{von Neumann for indis}\\
&=& S_{\text{Sh}}(0, \theta )-1. 
\label{von Neumann and Shannon relation}
\end{eqnarray}
This is different from the von Neumann entropy obtained from Eq. (\ref{superposed spin-singlet in the frame A with 0 and theta}); Eq. (\ref{von Neumann for indis}) is symmetric with respect to $\theta =\frac{\pi}{2}$. On the other hand, the Shannon entropy obtained from Eq. (\ref{superposed spin-singlet in the frame A with 0 and theta}) and from Eq. (\ref{no-distinguished particles state}) are found to be identical.

\subsection{The extreme value of the Shannon entropy and the von Neumann entropy of indistinguishable state}

In the von Neumann entropy for the distinguishable state, $\alpha$ and $\phi$ satisfy the relation Eq. (\ref{alpha_phi_relation}) at the maximum value and it is given by Eq. (\ref{the extreme value of distinguishable state}). On the other hand, in the Shannon entropy and the von Neumann entropy for the indistinguishable state, the following relation is satisfied along the extreme value 
\begin{eqnarray}
\left(
\frac{\cosh \alpha + \cosh \phi }{\sinh \alpha \sinh \phi}
\right)^2
=
\frac{(p_1-p_2)\sin ^2 \theta + \sqrt{(p_1-p_2)^2 \sin ^4 \theta +4 \cos ^2 \theta}}{2}. 
\label{alpha_phi_relation2}
\end{eqnarray}
This relation depends not only on $\theta$ but also on $p_1$ and $p_2$ unlike the relation (\ref{alpha_phi_relation}).

When $p_1-p_2 > 0$, S has a finite value for any $\theta$ and the extreme value is $2$ in the Shannon entropy or $1$ in the von Neumann entropy as the maximum value. Next, when $p_1-p_2 < 0$, the right hand of Eq. (\ref{alpha_phi_relation2}) is $0$ for $\theta = \frac{\pi}{2}$; this implies the super-relativistic limit. In this limit, the extreme value is
\begin{eqnarray}
S_{\text{Sh(ex)}} (0, \frac{\pi }{2} ) = -p_1 \log p_1 - p_2 \log p_2 + 1, \label{the extreme value of Shannon in pi/2}
\end{eqnarray}
and
\begin{eqnarray}
S_{\text{vN(ex)}} (0, \frac{\pi }{2} ) &=& -p_1 \log p_1 - p_2 \log p_2, \label{the extreme value of indistinguishable von Neumann in pi / 2}
\end{eqnarray}
since $|a_{\Lambda p}(0)|^2=|b_{\Lambda p}(\frac{\pi}{2})|^2=0$ and $|b_{\Lambda p}(0)|^2=|a_{\Lambda p}(\frac{\pi}{2})|^2=1$. For $\theta \ne \frac{\pi }{2}$, the extreme value is $2$ for the Shannon entropy and $1$ for the von Neumann entropy as the maximum value. 

In $\theta = \frac{\pi}{2}$, since $b_{\Lambda p}(\frac{\pi }{2})=0$ and then the state for $M_{\Lambda p}(\frac{\pi }{2})$ is the singlet-state, $p_2$ is the weight of the singlet-state. But, $p_1$ is the weight of the mixed state of the singlet-state and the triplet-state as seen in Eq. (\ref{the reduced density matrix of indistinguishable state}). Therefore, when $\theta = \frac{\pi}{2}$ and $p_1-p_2 < 0$, the singlet-state weight is always greater than the triplet-state weight; and then, the extreme value is less than the maximum value of $2$ or $1$, because the mixing of the singlet-state and the triplet-state does not become equal. In the super-relativistic limit, the state is a mixture of the singlet-state with the probability $p_2$ and the triplet-state with the probability $p_1$. On the other hand, when $\theta = \frac{\pi}{2}$ and $p_1-p_2 > 0$, the singlet-state weight and the triplet-state weight is equal at the point satisfying the relation (\ref{alpha_phi_relation2}) and then the extreme value takes the maximum value of $2$ or $1$.

When $p_1=p_2=\frac{1}{2}$, Eq. (\ref{alpha_phi_relation2}) equals to Eq. (\ref{alpha_phi_relation}). Then, in $\theta \geq \frac{\pi}{2}$, the von Neumann entropy for the distinguishable state, the Shannon entropy and the von Neumann entropy for the indistinguishable state behave similarly. In this special case, various features degenerate in $\theta \geq \frac{\pi}{2}$. 

Finally, We can also see, if $p_1 \gg p_2$ or $p_1 \ll p_2$, the von Neumann entropy of distinguishable state converges to 0; in other words, the state represented by $\rho '_{\Lambda }(0, \theta)$ tends to a pure state. But the Shannon entropy and the von Neumann entropy of indistinguishable state take the maximum value of 2 or 1 respectively. 

\section{Discussion}\label{Discussion}

The spin is known to be relativistically non-covariant concept and a covariant quantity such as the Pauli-Lubanski vector constructed from spin and 4-momentum has been proposed. However, in our discussion of entropy, the spin and the 4-momentum are treated independently, such as taking the trace over 4-momentum to get the reduced density matrix. Thus, the entropy is not relativistically covariant quantity. 

In our specific model, the ejection velocity of each particle and the boost velocity are both restricted on the $x$-$z$ plane and the spin state is assigned with respect to the spatially-fixed $z$-direction in the laboratory frame as well as in the center of mass frame. For the two pair states case, the state vector in the laboratory frame becomes an entangled state between 4-momentum part and spin part. The von Neumann entropy which is calculated from the reduced density matrix obtained by taking the trace over 4-momentum is dependent on the velocity parameters $\alpha $ and $\phi$, as well as $\theta$. This result is explained by the fact that the spin direction in the laboratory frame is inclined by the Wigner rotation relative to the rest frame and the spin correlation is described through $a_{\Lambda p}(0)$, $b_{\Lambda p}(0)$, $a_{\Lambda p}(\theta)$ and $b_{\Lambda p}(\theta)$ as seen in Eq. (\ref{the reduced density matrix 0 and theta}). 

In our specific model of two pair states with $\theta_1=0$ and $\theta_2=\theta$, the Shannon entropy is symmetric with respect to $\theta = \frac{\pi }{2}$ but the von Neumann entropy is not symmetric. 
This asymmetry comes from the distinguishability of two particles. 
The Shannon entropy is regarded to be the von Neumann entropy derived from the following density matrix $\rho =\frac{1}{2} (a|\uparrow \downarrow \rangle \langle \uparrow \downarrow |+b|\uparrow\uparrow \rangle \langle \uparrow\uparrow |+b|\downarrow\downarrow \rangle \langle \downarrow\downarrow |+a|\downarrow \uparrow \rangle \langle \downarrow \uparrow |)$, where $a=p_1 |a_{\Lambda p}(0)|^2+ p_2 |a_{\Lambda p}(\theta )|^2$ and $b=p_1 |b_{\Lambda p}(0)|^2+ p_2 |b_{\Lambda p}(\theta )|^2$. This density matrix shows the state of the measured system. 

Under the assumption of distinguishability, $|\varphi _p(\theta) \rangle $ is not equal to $|\varphi _p (\theta+\pi) \rangle $ but each state becomes identical after we have taken the trace over 4-momentum. $U(\Lambda )| \varphi _p (\theta) \rangle $ and $U(\Lambda )| \varphi _p (\theta + \pi) \rangle $ are a factor state of 4-momentum part and spin part. However, if we prepare the indistinguishable state as $|\psi \rangle = \frac{1}{\sqrt{2}}(|\varphi _p (\theta)\rangle - |\varphi _p (\theta + \pi) \rangle )$, the indistinguishable state $U(\Lambda ) |\psi \rangle $ is no longer a factor state and then the von Neumann entropy has a finite value. When $|\psi \rangle = \frac{1}{\sqrt{2}}(|\varphi _p (0)\rangle - |\varphi _p (\pi) \rangle )$, the von Neumann entropy $S(\phi,\alpha ,0,\pi)$ takes the same form with Eq. (\ref{the von Neumann entropy of distinguishable state 0 and theta}). Therefore, we may regard the indistinguishable state for one pair state is a special case of two pair states. Thus, even if $p_1 \gg p_2$ or $p_1 \ll  p_2$, the von Neumann entropy for the indistinguishable state has the maximum value of 1, where the singlet-state and the triplet-state are equally mixed.

Under the assumption of indistinguishability, the von Neumann entropy coincides with the Shannon entropy, except for the difference of absolute value by 1 as seen in Eq. (\ref{von Neumann and Shannon relation}). However, if we want only to see the effect of the Lorentz transformation on the entropy, the Shannon entropy should be renormalized to take 0, instead of the value of 1, for the state in $\alpha=0 $ and $\phi=0$. 

\section*{Acknowledgements}

The author would like to express his sincere gratitude to Professor Humitaka Sato for his very important suggestions, useful discussions, his valuable advice and continuous support throughout this work. The author would like to thank Professor Masahisa Ohta for his various supports and encouragement. The author also would like to thank Professor Hajime Susa and Professor Takahiro Wada for their warm encouragement and support. The author wishes to thank his family and all the member of the theoretical physics group of Konan University for kind support.

%

\end{document}